\documentclass[prl,aps,showpacs]{revtex4}
\usepackage{ifpdf}
\usepackage{amsmath}
\usepackage{graphicx}
	\usepackage{amssymb}

\begin{document}
%\date{28.6.2005}

\title{The fate of black branes in Einstein-Gauss-Bonnet gravity}
\author{P. Suranyi$^a$, C. Vaz$^{a,b}$, and L.C.R. Wijewardhana$^a$}
\affiliation{${}^a$Department of Physics, University of Cincinnati, Cincinnati, Ohio, 45221-0011\\ ${}^b$Raymond Walters College, University of Cincinnati, Cincinnati, Ohio, 45221}

\begin{abstract} Black branes are studied in Einstein-Gauss-Bonnet (EGB) gravity.  Evaporation drives black branes  towards one of two singularities depending on the sign of $\alpha$, the Gauss-Bonnet coupling.  For positive $\alpha$ and sufficiently large ratio $\sqrt{\alpha}/L$, where $L/2\pi$ is the  radius of compactification, black branes avoid the Gregory-Laflamme (GL) instability before reaching a critical state. No black branes with the radius of horizon smaller than the critical value can exist. Approaching the critical state branes have a nonzero Hawking temperature. For negative $\alpha$ all black branes encounter the GL instability. No black branes may exist outside of the interval of the critical values, $0\leq\beta<3$, where $\beta=1- 8\alpha/r_h^2$ and $r_h$ is the radius of horizon of the black brane. The first order phase transition line of GL transitions ends in a second order phase transition point at $\beta=0$. 
 \end{abstract}
 \maketitle
\section{Introduction}
Gravity in higher dimensions has been in the forefront of research for a considerable time.  Most of the studies start with the conventional Einstein action generalized to $D>4$ dimensions. In the low energy limit quantum theories of gravity like string theory yield additional higher order curvature correction terms to the Einstein action.\cite{callen} Some higher order curvature terms when taken in isolation and quantized can potentially lead to problematic high energy behavior, such as the presence of negative norm states. The simplest term avoiding these problems is the Gauss-Bonnet  (GB) term.  In this letter we analyze the critical behavior of a class of classical solutions to Einstein gravity, augmented by the GB term.  Such a term arises in the gravity action in string theory when corrections to zero slope limit are calculated~\cite{nepomichie}~\cite{boulware}. For simplicity, we restrict our discussion to five space-time dimension, but as we will point out later the results can be simply generalized to higher dimensions. 

In Einstein gravity $D=4$ solutions, like a black hole can be trivially extended to a black string ($D=5$) or black brane solutions ($D>5$). Solutions become more complicated if higher derivative terms are added to the Einstein action, such as the GB term. Though asymptotically Minkowski black {\em hole} solutions have been found in $D\geq5$ theories~\cite{boulware}, no exact black string (brane) solutions are known in EGB gravity. Yet the investigations of black strings (branes) is more important because with compact extra dimensions all but the lightest static objects must be black strings (branes). 

Two alternative techniques have been used to investigate black brane solutions. Kobayashi and Tanaka~\cite{kobayashi} solved the 5 dimensional Einstein equations modified by the Lanczos tensor (variation of the Gauss-Bonnet term), numerically.   Using an expansion around the event horizon they also found an exact lower bound for the black string mass, $M_{c}\sim \sqrt{\alpha}$, where $\alpha$ is the coupling constant of the GB term. For $M<M_{c}$ the solutions do not have a horizon, they represent naked singularities. 

In a subsequent work~\cite{chetiya}, following~\cite{kobayashi}, we used a series expansion in $\alpha$ to investigate black strings in $D>5$ EGB theory.  We found that in every dimension, $D\geq 5$, a lower bound, similar to that in $D=5$, exists for the mass of the black string.  We also investigated the thermodynamics of the black string but the fifth order expansion employed in our paper was insufficient to get a definitive answer for the thermal behavior of the system when the mass of the black string approached the lower limit. 

In this paper we also use improved numerical and expansion methods to investigate black string solutions including the neighborhood of critical points, $\beta_{\rm crit}=0$ and 3, where the dimensionless GB coupling is defined as 
\begin{equation}
\beta=1- 8\frac{\alpha}{r_h^2}.
\label{beta}
\end{equation}
In (\ref{beta}) $r_h$ denotes the radius of the horizon. $\beta =0$ determines the critical radius. The critical radius has a one-to-one relationship with the critical mass, which is determined by the condition $\beta>0$.

In Section 2 we will discuss the uniform black string, which should have a higher entropy than a black hole, when the mass is large enough.  In fact, for sufficiently high mass it is the only static state of the system.  We will use a horizon expansion to elucidate the nature of two critical points  $\beta=0$ and 3.  We will concentrate on $D=5$ but, except for numerical details $D\geq 6$ theories are not different.  Then we will use numerical calculations to investigate the properties of black string at large distances from the horizon.  We discuss their thermodynamics and calculate their entropy. 

In Section 3 we will investigate the extension of the Gregory-Laflamme instability~\cite{gregory} and the notion of the non-uniform black string, which is the  preferred state if the radius of horizon goes under a critical multiple of the radius of compactification~\cite{wiseman}~\cite{gubser}~\cite{harmark}.  We will also plot the phase diagram of black strings in the $\beta-r_h$ plane.

Section 4. summarizes our results. We will also discuss the possible directions for the continuation of this research.

\section{Uniform black string}

 In this section we will discuss the extension of a 4-dimensional solution to $D$.  If we assume that the metric $g_{ab}$ is an asymptotically flat solution of the 4 dimensional Einstein equation then the trivial extension of this metric, obtained by defining $g_{ai}=0$ and $g_{ii}=1$ satisfies the $D$ dimensional Einstein equation.  Here, and in what follows, $D\geq i\geq5$. We tacitly assume that the coordinates, labeled by $i$ are compactified on circles, in which space we will maintain a $(S^1)^{D-4}$ symmetry throughout this paper.  However, the trivially extended metric is not the solution of Einstein-Gauss-Bonnet equation
\begin{equation}
G_{ab}\equiv R_{ab}-\frac{1}{2}g_{ab}R+\alpha L_{ab}=0,
\label{egb}
\end{equation}
where $\alpha$ is the coupling, the Lanczos tensor is
\begin{equation}
L_{{ab}}=-\frac{1}{2}g_{ab}\left(R^{2}-4R^{ab}R_{ab}+K\right)+2RR_{ab}-4R_{ac}R_{b}^{c}-2R_{acde}R_{b}^{cde}+4R_{acdb}R^{cd}
\label{lanczos}
\end{equation}
and
$K$ is the Kretschmann scalar,
\begin{equation}
K=R_{{abcd}}R^{abcd}.
\label{kretschmann}
\end{equation}

In an appropriate coordinate system the metric components of uniform black brane solutions depend only on a radial coordinate.  Nonuniform black branes and the Gregory-Laflamme instability will be discussed later. The metric of nonuniform black branes is a periodic function of the compactified coordinate, $w$, as well.

\subsection{Horizon expansion and scaling near critical points}

Throughout this paper we will use dimensionless coordinates, scaled by the radius of the horizon, $r_h$, of the uniform black string. In particular, we choose the horizon variable, $\rho$ defined by
\begin{equation}
\rho=\frac{r-r_h}{r_h},
\label{rho}
\end{equation}
to replace the radial coordinate, $r$.  When the GB coupling, $\alpha$, is also rescaled, by the introduction of the  parameter, $\beta$ (See eq. \ref{beta}), then all reference to the radius of the horizon is eliminated from the equations of motion.  The only parameter remaining in the equations of motion is $\beta$.  True values of physical quantities are restored by multiplying by a power of  $r_h$, required by their dimensions.  

Defining the metric of a uniform string, made dimensionless by  factoring out $r_h^2$, as 
\begin{equation}
ds^2=-f(\rho)dt^2+\frac{g(\rho)}{f(\rho)}d\rho^2+(\rho+1)^2d\Omega^2+h(\rho)dw^2,
\label{metric}
\end{equation}
we can solve the equations of motion in a power series of $\rho$ (horizon expansion).  The series coefficients depend only on $\beta$. 
Displaying three leading orders only, the expansions have the following form:
\begin{eqnarray}
f(\rho)&=& f_1\,\left( \rho -\frac{-\beta +\sqrt{\beta }+15-\frac{7}{\sqrt{\beta }}}{4(3- \beta )}\rho^2-\frac{-55 \beta ^3-76 \beta ^{5/2}+61 \beta ^2-416 \beta ^{3/2}+135 \beta
   -84 \sqrt{\beta }+147}{72 (\beta -3)^2 \beta ^{3/2}}\rho^3+...\right)\nonumber\\
   g(\rho)&=&\frac{\sqrt{\beta }+1}{2}+\frac{7-8\beta +\beta^2}{4(3- \beta)\sqrt{\beta} }\rho\nonumber\\
   &+&\frac{\left(1+\sqrt{\beta}\right) \left(-147+231\sqrt{\beta}-261\beta+161\beta^{3/2}+47\beta^2-35\beta^{5/2}+\beta^3+3\beta^{7/2}
  \right)}{72 (\beta -3)^2 \beta ^{3/2}}\rho^2+...\nonumber\\
h(\rho)&=&h_0\left(1-\frac{4 \left(\sqrt{\beta }-1\right)}{\beta -3}\rho+\frac{2 \left(\beta ^2+4 \beta ^{3/2}-16 \beta +4 \sqrt{\beta
   }+7\right)}{(\beta -3)^2 \sqrt{\beta }}\rho^2+..\right)
   \label{horizon1}
   \end{eqnarray}
The two constants, $f_1$ and $h_0$, are not determined by the horizon expansion alone.  They  depend on $\beta$ only. They must be introduced in $f(\rho)$ and $h(\rho)$ to assure that the metric components are asymptotically Minkowski. They are determined in the process of numerical integration of the equations of motion.

The expansion coefficients have two singular points: $\beta=0$ and $\beta=3$. Note that for  $\alpha<0$ $\beta>1$. The physical domain is $0\leq \beta< 3$.  The expansion coefficients are complex for $\beta<0$ and there are singularities outside the horizon for $\beta\geq3$. Questions have been raised whether negative values of $\alpha$ are admissible ~\cite{arkani}. Nevertheless, for completeness we will also investigate that region, as well. Near the singular points the convergence radius of the expansion shrinks to zero.  The leading singular terms of the expansion coefficients (subscript $k$ implies the coefficient of $\rho^k$) are the following:
\begin{eqnarray}
f_k&\sim&   \beta^{3/2-k},\,\,\, (3-\beta)^{1-k},\nonumber\\
g_k&\sim& \beta^{1/2-k},\,\,\, (3-\beta)^{-k},\nonumber\\
h_k&\sim&  \beta^{3/2-k}.\,\,\, (3-\beta)^{-k}.
\end{eqnarray}

Since the expansion coefficients become complex at $\beta<0$ the solution does not exist for negative $\beta$, contradicting the assumption of  the existence of a horizon or/and a regular expansion around the horizon~\cite{kobayashi}. This result can, however, be the result of a poor choice of the coordinates. Therefore,  we also calculate scalars, like the Ricci scalar ($R$) and the Kretschmann scalar ($K$), at the horizon.  Being scalars, they must be independent of the scaling factors $f_1$ or $h_0$ which can be gauged away by rescaling the coordinates $t$ or $w$.  Indeed, we obtain
\begin{eqnarray}
R&=&-2\frac{1+\sqrt{\beta}-3\beta+\beta^{3/2}}{(3-\beta)(1+\sqrt{\beta})},\nonumber\\
K&=&4\frac{69+2\sqrt{\beta}-13\beta-12\beta^{3/2}-\beta^2+2\beta^{5/2}+\beta^3}{(3-\beta)^2(1+\sqrt{\beta})^2}
\label{scalars}
\end{eqnarray}
We can draw several conclusions from (\ref{scalars}). First of all, $R$ and $K$ are non-analytic at $\beta=0$ and 3, ruling out mere coordinate singularities. Furthermore, both of these constants become complex for $\beta<0$, therefore no uniform black string solution with regular horizon can exist in that region. $\beta<0$ implies $r_h<\sqrt{8\alpha}$ and, as we will see later, $r_h$, though not quite proportional, but is a regular and bounded function of the ADM mass. Thus, the $\beta>0$ bound imposes a lower limit on the mass of the black string.  

Furthermore, while all curvature invariants are finite at $\beta=0$ they are infinite at $\beta=3$. This points to the possibility  of the existence of a critical solution at $\beta=0$ but not at $\beta=3$. We will discuss the $\beta=0$ critical solution in the following section.

\subsection{The critical string}

To investigate the critical solution we start with a rearrangement of the horizon expansion series. Keeping only the leading singular terms in every order of $\rho$ we are lead to  functions of a scaling variable $y=\rho/\beta$. Thus, one can define alternative expansions in series of scaling functions $\phi_i(y),\gamma_i(y),$ 
and $\eta_i(y)$. The scaling expansion is defined as
\begin{equation}
f(\rho)=\rho\left \{1+\frac{\rho}{\sqrt{\beta}}[\phi_1(y)+ \sqrt{\beta}\phi_2(y)+...]\right\},
\label{expansion2}
\end{equation}
with similar expressions for $g(\rho)$ and $h(\rho)$ in terms of series of scaling functions, $\gamma_i(y)$ and $\eta_i(y)$.  Substituting (\ref{expansion2}) into (\ref{egb}) we obtain differential equations for the scaling functions defined in (\ref{expansion2}). 
The equation for $\phi_1$ is
\begin{equation}
[6y^3(2y\phi_1-1)+6y^5\phi_1']\phi_1''+27y^4(\phi_1')^2+6y^2(12y\phi_1-5)\phi_1'+2(18y2\phi_1^2-12y\phi_1-7y)=0,
\label{phi1eq}
   \end{equation}
while $\gamma_1$ and $\eta_1$ can be simply expressed by $\phi_1$ as   $\gamma_1= 2\phi_1+y\phi_1'$ and
$\eta_1=4 \phi_1/3$.
 Though we could not find the solution of (\ref{phi1eq}) in analytic form the general behavior of $\phi_1$ can be analyzed fairly easily.  Its power series in $y$ is provided by the most singular terms of horizon expansion (\ref{horizon1}).  Analyzing an 80th order series expansion we were able to ascertain that $\phi_1$ has a  singularity inside the horizon of the form
 \begin{equation}
 \phi_1\sim \left(y+\frac{3}{7}\right)^{3/2}.
 \label{singularity}
 \end{equation}
Eq.  (\ref{singularity}) suggests that the metric becomes singular inside the horizon, at $\rho=-3 \beta/7$ and cannot be continued to smaller values of $\rho$. Note that this singularity moves to $\rho=0$ at $\beta=0$. We will indeed see below that the critical solution is singular at $\rho=0$ and cannot be continued to $\rho<0$. There is no impediment, however to continue $\phi_1$ to positive values. In fact the analysis of Eq. (\ref{phi1eq}) shows that the asymptotic behavior of $\phi_1$ is 
 \begin{equation}
 \phi_1\simeq \pm\frac{2}{3}\sqrt{\frac{7}{3}}y^{-1/2}+O(y^{-1}).
 \label{asymptotic3}
 \end{equation}
Numerical integration of Eq. (\ref{phi1eq}) shows that for a function with the series expansion implied by (\ref{horizon1})  the positive sign should be chosen in (\ref{asymptotic3}). Now in the $\beta\to0$ limit the first terms of  (\ref{expansion2}) become
\begin{equation}
f(\rho)=\rho\left \{1+\frac{\rho}{\sqrt{\beta}}\left[\frac{2}{3}\sqrt{\frac{7}{3}}\sqrt{\frac{\beta}{\rho}}+...\right]\right\}=\rho\left \{1+\sqrt{\rho}\frac{2}{3}\sqrt{\frac{7}{3}}+...]\right\}
\label{expansion3}
\end{equation}
The $\beta\to0$ limit of higher order expansion terms lead to higher powers of $\rho$, including all half-integer powers.  This can be ascertained by repeating the horizon expansion (\ref{horizon1}) with half-integer powers included (starting from $\rho^{3/2}$). This leads to the following expansion of the critical black string ($\beta=0$) solution:
\begin{eqnarray}
f(\rho)&=&f_1\left(\rho+\frac{2}{3}\sqrt{\frac{7}{3}}\rho^{3/2}-\frac{67}{90}\rho^2-\frac{22853}{6750\sqrt{21}}\rho^{5/2}+\frac{250322}{893025}\rho^3+...\right),\nonumber\\
g(\rho)&=&\frac{1}{2}+\sqrt{\frac{7}{3}}\sqrt{\rho}+\frac{91}{90}\rho-\frac{8993}{2600\sqrt{21}}\rho^{3/2}-\frac{1055977}{1190700}\rho^2+...,\nonumber\\
h(\rho)&=&h_0\left(1-\frac{4}{3}\rho+ \frac{8}{9}\sqrt{\frac{7}{3}}\rho^{3/2} + \frac{46}{35} \rho^2+...\right).
\label{critical}
\end{eqnarray}
If one attempts an expansion with half-integer powers of $\rho$ for general $\beta$ then either all the half-integer powers vanish or $\beta=0$. Conversely, no horizon expansion with integer powers exists for the critical black string.  

The value of scalars $R$ and $K$ can also be calculated at $\rho=0$ using (\ref{critical}). One obtains $R=-2/3$ and $K=92/3$, precisely the values obtained from the $\beta=0$ limit of (\ref{scalars}). In fact, one can show that all scalars are finite in the critical state.  Correction to scalars, when one moves away from the horizon are of $O(\sqrt{\rho})$. Further evidence for the fact that  the critical string is the limit of non-critical strings when $\beta\to0$ comes from numerical integration of the equations of motion, to be discussed later.

The $\rho=0$ surface cannot really be called a horizon, just the limit of horizons when $\beta\to0$. While it is an infinite red shift null surface, the metric cannot be continued inside the $\rho=0$ surface. We believe that there is  no way to facilitate such a continuation.  Strictly speaking, though this is a  string solution, it is not "black."  It is however a limiting solution of a series of black string solutions in every quantity that we investigated.  At all $\beta>0$ the horizon is completely regular and there is always a region inside the horizon, $\rho>-3\beta/7$ , where the solution can be continued without any impediment.  

Time-like geodesics can be continued to the horizon both in the $\beta>0$ and $\beta=0$ cases. A test particle reaches the horizon in finite proper time, $\tau$.  However, while in the $\beta>0$ case the test particle passes through the horizon into a region from which it cannot escape any more, for $\beta=0$ geodesics have a pathological behavior. Choosing a geodesic for a particle falling towards the horizon, $\dot\rho(\tau)<0$, and the time scale such that  $\rho(0)=0$ the radial coordinate turns complex for $\tau>0$.  Therefore, the role of the critical solution as a valid state of black strings is questionable. This question will be discussed later in more details, following the discussion of numerical integration of the equations of motion and of the Gregory-Laflamme instability.

\subsection{Numerical integration of the equations of motion}

We build on the results of the previous sections to obtain the metric functions at arbitrary $\rho>0$. Before starting the numerical integration from the horizon we calculate the value of the metric functions and their derivatives from the horizon expansion at $\rho=\Delta>0$. This  step is necessary because $\rho=0$ is a singular point of the differential equations.  $\Delta$ is chosen to be small enough so that the last terms of the 30th order horizon expansion is smaller than $10^{-10}$.  Then we integrate the equations numerically from $\rho=\Delta$ using metric (\ref{metric}).   The step size, $\delta$, in the numerical integration is chosen such that $\delta<0.01\Delta$.  

Expansions (\ref{horizon1}) contain undetermined constants $f_1$ and $h_0$. In an initial run these constants are chosen to be 1. Then, though the calculated metric functions tend to a finite value as $\rho\to\infty$, their asymptotic limit is not  1.  Fitting a form 
\begin{equation}
f(\rho)= a_f+ \frac{b_f}{\rho}+\frac{c_f}{\rho^2}+...,
\label{asymptotic}
\end{equation}
to $f(\rho)$, and a similar form to $h(\rho)$ we define $f_1=1/a_f$ and $h_0=1/a_h$ to get, after a second round of numerical integrations, metric functions that are asymptotically 1, which corresponds to Minkowski space. 

\subsection{Thermodynamics}

The numerical form of the metric functions allows us to investigate the thermodynamic properties of the black string.  The first law of thermodynamics states that ~\cite{gubser}\cite{harmark}
\begin{equation}
dS= \frac{dM}{T}- \frac{\kappa \, dL}{T},
\end{equation}
where $M$ is the Arnowit-Deser-Misner (ADM) mass, $T$ is the Hawking temperature of the black string, $\kappa$ is the tension, and $L$ is circumference of the compact string.

The surface gravity on the horizon, $k$,  is related to the Hawking temperature as follows ~\cite{wald}~\cite{myers}~\cite{solodukhin}
\begin{equation}
T=\frac{k}{2\pi}.
\label{temp}
\end{equation}
  Using (\ref{metric}) and (\ref{horizon1}) we obtain
\begin{equation}
k=\frac{1}{r_h}\frac{f'(0)}{2\sqrt{g(0)}},
\label{surfacegravity}
\end{equation}
where we restored the correct dimension of the surface gravity by reinstating a factor of $r_h^{-1}$.
As we pointed out earlier $f'(0)$ and $g(0)$ are determined in the process of numerical integration. They are functions of $\beta$ only. Combining (\ref{temp}) and (\ref{surfacegravity}) we can write the Hawking temperature as
\begin{equation}
T= \frac{1}{4\pi r_h}F_T(\beta).
\label{hawk1}
\end{equation}
Form factor 
$F_T(\beta)$, normalized to $F_T(1)=1$, is calculated in the process of our numerical integration procedure. It is a smooth function of $\beta$ between $\beta=0$ and $\beta=3$. It tends to a finite value at $\beta=0$. That limit, along with the limits of the other two form factors defined below, agree with the values obtained from the critical string at $\beta=0$.  $F_T(\beta)\to\infty$ in the limit $\beta\to3$  implying that $T\to\infty$ as $\beta\to3$. $F_T$ and the other two form factors are plotted in Fig.1, 

The ADM mass and the tension are expressed simply by the long range components of $f(r)$ and $h(r)$~\cite{harmark}. Namely, after renormalizing $f(\rho)$ and $h(\rho)$ so that $f(\infty)=1, \,h(\infty)=1$  and defining  the leading, $O(\rho^{-1})$, asymptotic coefficients as $f_{-1}$ and $h_{-1}$, then 
\begin{eqnarray}
M&=&\frac{L r_h}{4G_5}(2f_{-1}-h_{-1})=\frac{L r_h}{2G_5}F_M(\beta)\label{m}\\
\tau&=&\frac{r_h}{4G_5}(f_{-1}-2h_{-1})=\frac{r_h}{4G_5}F_\tau(\beta),\label{tau}
\end{eqnarray}
where 
$F_M(\beta)$ and $F_\tau(\beta)$ are two additional form factors depending on $\beta$ only.   They are both bounded on the interval $0\leq\beta\leq3$. These form factors also converge smoothly to the corresponding values for the critical string at $\beta=0$. They are plotted in Fig 1. along with $F_T(\beta)$. All three of the form factors are normalized to  1 at $\beta=1$ (vanishing GB coupling).
\begin{figure}[htbp]
\begin{center}
\includegraphics[width=4in]{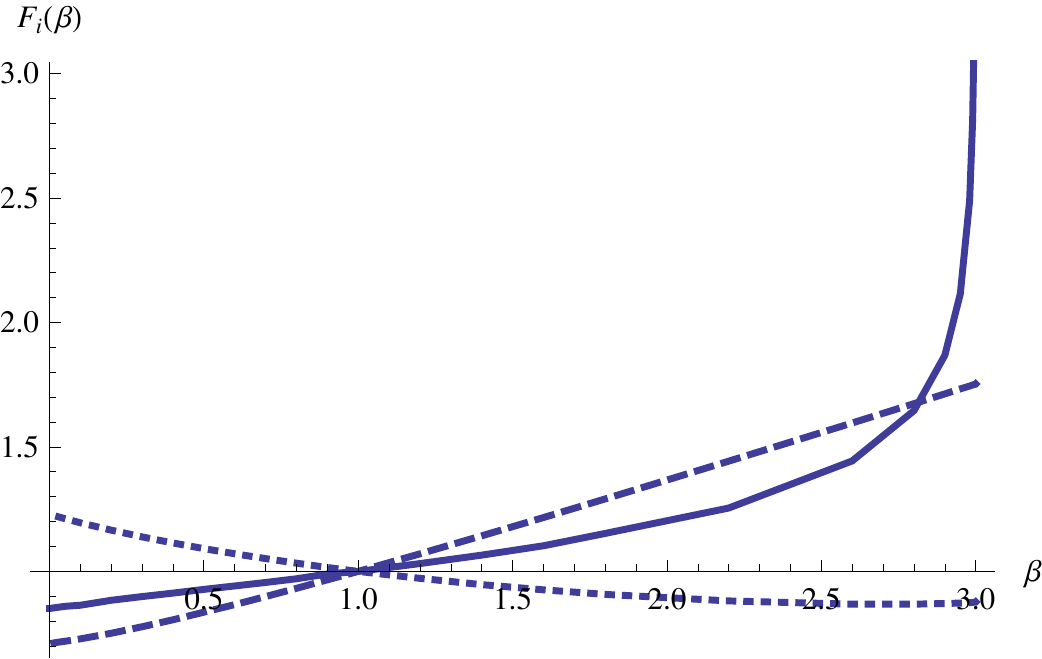}
\caption{The plot of the scaling functions $F_T(\beta)$ (solid line), $F_M(\beta)$ (dotted line) and $F_\tau(\beta)$ (dashed line) as a function of $\beta$ between the two critical points, 0 and 3} 
\label{Fig.1}
\end{center}
\end{figure}
The three form factors are not independent from each other.  The integrability condition
\begin{equation}
\frac{\partial T^{-1}}{\partial L}+\frac{\partial(\tau\,T^{-1})}{\partial M}=0.
\label{integrability}
\end{equation}
relates them. (\ref{integrability}) implies
\begin{equation}
(1-\beta ) (2 F_M-F_\tau) F_T'+F_T \left(-F_M+F_\tau+(1-\beta )F_\tau'\right)=0.
   \label{relationship}
   \end{equation}
   The calculated values of the form factors satisfy (\ref{relationship}) to $O(10^{-4})$, except near $\beta=3$, where the rapid variation of $F_T$ makes the evaluation of the numerical derivative difficult.
   
   We can also calculate the entropy.  We have
   \begin{equation}
   S=\int \frac{dM}{T}+\frac{8\pi \alpha}{G}q(L)
   \label{entropy3}
   \end{equation}
   where $q(L)$ is independent of $M$. (\ref{entropy3}) can be rewritten by changing variables as
   \begin{equation}
S= \frac{8 L \pi  \alpha}{G}  \int \frac{F_M+2 (1-\beta ) F_M'}{(1-\beta )^2 F_T} \, d\beta +  \frac{8 \pi  \alpha  }{G}q(L)
\label{entropy4}
   \end{equation}
   The function $q(L)$, a yet undetermined arbitrary function of $L$, can be fixed if we take the derivative of (\ref{entropy4}) with respect to $L$ and use the integrability condition (\ref{relationship}). Setting the lower bound of the integration at $\beta=0$ we obtain the final form of the entropy for $\alpha>0$ as 
    \begin{equation}
S= \frac{8 L \pi  \alpha}{G} \left( \int_0^{1-8\alpha/r_h^2} \frac{F_M(\beta)+2 (1-\beta ) F_M'(\beta)}{(1-\beta )^2 F_T(\beta)} \, d\beta +   \frac{2F_M(0)-F_\tau(0)}{F_T(0)}\right)
\label{entropy5}
   \end{equation}
An overall constant was fixed in (\ref{entropy5}) so that $S=0$ for $L=0$.  From our numerical integration we obtain $[2F_M(0)-F_\tau(0)]/F_T(0)=2.05614$.
   
For completeness we write down an equivalent expression for $\alpha<0$.  Considering that $F_T$ diverges around the critical point $\beta=3$ the integral over $\beta$ is convergent at that point so we can write
  \begin{equation}
S=- \frac{8 L \pi  \alpha}{G} \int_{1-8\alpha/r_h^2}^3 \frac{F_M(\beta)+2 (1-\beta ) F_M'(\beta)}{(1-\beta )^2 F_T(\beta)} \, d\beta,
\label{entropy5b}
\end{equation}
implying that the entropy vanishes at the critical point, $\beta=3$.

The limit of $S$ when $\alpha\to 0$ can be calculated if we change the integration variable in (\ref{entropy5}) to $r_h$. Then we obtain
 \begin{equation}
S= \frac{2 L \pi  }{G} \left( \int_{\sqrt{8\alpha}}^{r_h} \frac{F_M(\beta(r_h'))+r_h' \partial F_M(\beta(r_h'))/\partial r_h'}{ F_T(\beta(r_h'))} r_h'\, dr_h'+   4\alpha\frac{2F_M(0)-F_\tau(0)}{F_T(0)}\right)
\label{entropy6}
   \end{equation}
Since $\beta(r_h')=1-8\alpha/r_h'^2$ and the form factors are equal to unity at $\beta=1$, at that point (\ref{entropy6}) reduces to
\begin{equation}
S_0= \frac{2 L \pi  }{G} \left( \int_{0}^{r_h} r_h'\, dr_h'\right)=\frac{ L \pi r_h^2 }{G} ,
\label{entropy7}
\end{equation}
which is the expression for the entropy of a black string in Einstein gravity.
 \begin{figure}[htbp]
\begin{center}
\includegraphics[width=4in]{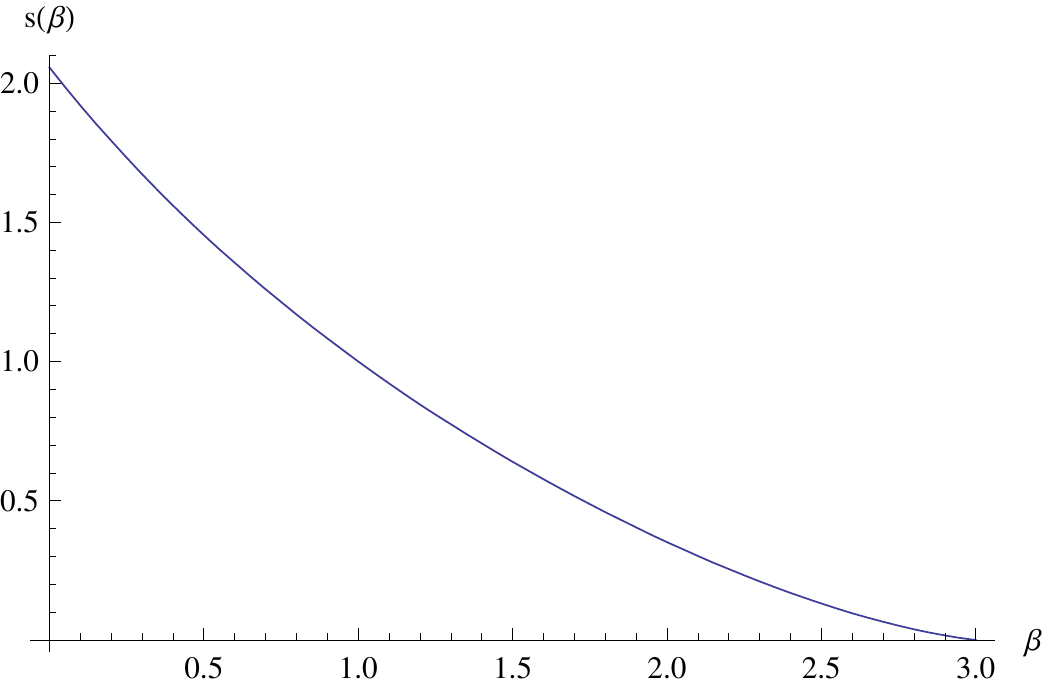}
\caption{Plot of function $s(\beta)$} 
\label{Fig.2}
\end{center}
\end{figure}
The entropy can be represented in the form
\begin{equation}
S= \frac{ L \pi r_h^2 }{G}s(\beta)
\label{entscale}
\end{equation}
at all $0\leq\beta<3$.  The function $s(\beta)$ obtained from numerical integration  is plotted in Fig.2. Note that $s(1)=1$ and $s(3)=0$.

\section{Gregory-Laflamme instability and nonuniform black strings}

\subsection{A simple picture of the life of a black string}

We will investigate what happens  to black strings  as Hawking radiation reduces their mass.  Consider the ADM mass as a function of $r_h$ and $\beta$, (\ref{m}).  $M$ is a monotonic function of $r_h$, because for the whole physical range of $\beta$
\begin{equation}
\frac{\partial M}{\partial r_h}=\frac{L}{2G_5}(F_M+2(1-\beta)F_M')>0.
\end{equation}
  With decreasing $r_h$ $\beta$ decreases if $\alpha>0$ and increases if $\alpha<0$. Typical trajectories of black strings in the $(\beta,2\pi r_h/L)$ plane are shown in Fig. 3.  According to (\ref{beta}) if $\alpha>0$  $r_h\geq\sqrt{8\alpha}$. When $r_h$ hits the value $\sqrt{8\alpha}$ the singular point, $\beta=0$, is reached. In the critical state the black string has a finite Hawking temperature, but it cannot decay and stay in the black string state any more.  We will return to the discussion of its further fate later. 
  
    If $\alpha<0$ $\beta$ increases with decreasing $r_h$. Eventually, it will hit the critical value $\beta=3$, where $r_h=\sqrt{-4\alpha}$. However, the Hawking temperature diverges at $\beta=3$, so approaching this point the black string will completely evaporate.  
    
    Before turning to the investigation of instability we briefly discuss the fate of black branes in $D>5$ EGB theories. We will use the notation $\bar\alpha=\alpha/ r_h^2$. The coefficients of the horizon expansion depend on $\bar\alpha$ only.  Their singularity structure in $\beta$ is the same as that at $D=5$, except $\beta$ has a more complicated expression
    \begin{equation*}
    \beta=1-48\bar\alpha^2-1152 \frac{D-4}{(D-2)^2}\bar\alpha^3.
    \end{equation*}
    $\beta$ has a single zero for $\alpha>0$, which zero varies with $D$ from a minimum of $\bar\alpha=0.1233$ at $D=6$ to $\bar\alpha=0.1443$ at $D=\infty$.  For negative $\alpha$ we have $\beta>0$ though $\beta$ is not a monotonic function of $\bar\alpha$. Still, just like for $D=5$ the horizon expansion coefficients are singular at $\bar\alpha=-1/4$.  Thus, aside for small numerical changes the fate of black branes is the same as that of black strings. They are driven to the critical state $\beta=0$ if $\alpha>0$ and to $\bar\alpha=-1/4$ if $\alpha<0$.
    
\subsection{Gregory-Laflamme instability}

In the absence of the GB interaction term ($\alpha=0$), uniform black strings become unstable  if $2\pi r_h/L<\kappa~\simeq0.876$~\cite{gregory}.  At this point a nonuniform mode becomes marginally tachyonic, a zero mode.   As it was shown by Gubser \cite{gubser}, Harmark and Obers,  Kol, Piran, and Sorkin~\cite{harmark}  this point corresponds to a first order transition to a stable nonuniform black string state.  Such a state is associated with a periodic dependence of the metric function on the compactified coordinates, in our case $w$.  It is fair to assume that such an instability also affects the EGB theory when $\alpha\not=0$. Therefore, the simplistic picture about the life of a black string which we depicted  in the previous section may be misleading. The black string may hit an instability line before it reaches either of the singular points ($\beta=0$ or 3).  Therefore, we performed a stability analysis of EGB black strings using our numerical approximation to the metric components of the uniform string solution. 

Following Refs. \cite{gubser} and \cite{harmark} we perturbed the metric with a periodic function with wave number $\kappa$ of of the compact coordinate, $w$.  Considering the rescaling of coordinates with $r_h$ $\kappa$ can be identified with $\kappa=2\pi r_h/L$. Applying the Landau-Ginzburg theory of phase transitions to the Einstein black string it was found  ~\cite{gubser}~\cite{harmark} that at the Gregory-Laflamme point~\cite{gregory}, $\kappa\simeq0.876$, the system undergoes a first order phase transition to a nonuniform state with a finite non-uniformity.  It is quite reasonable to expect that a first order transition exists at nonzero values of the GB coupling, as well. 

We consider the following static perturbation of metric (\ref{metric}) \cite{gubser} \cite{harmark}
\begin{equation}
ds^2=-e^{2A}f(\rho)dt^2+e^{2B}\frac{g(\rho)}{f(\rho)}d\rho^2+(\rho+1)^2e^{2C}d\Omega^2+e^{2H}h(\rho)dw^2,
\label{metric2}
\end{equation}
where the exponents $A,\,B,\,C,$ and $H$ are functions of both $\rho$ and $w$.  Though in (\ref{metric2}) we gauged away a possible non-diagonal term still the gauge has not been fixed completely.  A perturbation analysis is applied to (\ref{metric2}), assuming a small deviation from the uniform string.  It is also assumed that the smallest possible wavelength, $\kappa$ dominates the perturbation. Higher order corrections, along with higher harmonics can be considered in a systematic perturbation analysis that we leave to future work.  Accordingly, we use the following ansatz for the perturbed metric
\begin{equation}
A=1+ \lambda \,a(\rho)\cos(\kappa w).
\label{ansatz}
\end{equation}
We also introduce amplitudes $b(\rho)$, $c(\rho)$ and $h(\rho)$ for the perturbative form of coefficients $B,\,C,$ and $H$.  Linearization  of the equations of motion in $\lambda$ leads to a set of equation for the amplitudes $a,\,b,\,c$ and $h$.  These equations still have a substantial gauge freedom.  As we prefer to work with a single amplitude, we use the Kol-Sorkin gauge \cite{harmark} to eliminate $b,\,c$, and $h$, using the gauge freedom.  

Making the final gauge choice $c(\rho)=-a(\rho)/2$, $G_{rw}=0$ can be solved for $b(\rho)$. Here and in what follows we use the notation $G_{\mu\nu}$ for the components of the equation of motion.  In the next step $G_{rr}=0$ and $G_{\theta\theta}=0$ are used to eliminate $h(\rho)$ and its derivatives. Then $G_{tt}=0$ and $G_{ww}=0$ provide identical second order differential equations of the form
\begin{equation}
W(\rho)a''+ X(\rho)a'+[Y(\rho)+\kappa^2 Z(\rho)]a=0
\label{aeq}
\end{equation}
 for $a(\rho)$. 
 
 The coefficients $W$, $X$, $Y$, and $Z$ are extremely complicated functions of the metric components $f,\,g,$ $h$ and their derivatives \footnote[1]{The form of the coefficient functions can be found in the 'Mathematica' file "Coefficient  functions" at the internet address: http://www.physics.uc.edu/suranyi/mathematicafiles. }. An important feature of $W$ is its proportionality to $f$. Consequently $W$ vanishes on the horizon making (\ref{aeq}) singular at $\rho=0$.   The ratios $X/W\sim \rho^{-1}$, $Z/W\sim 1$ and $Y/W\sim \rho^{-3}$ when $\rho\to\infty$.  Thus, (\ref{aeq}) is a Schrodinger-like equation, with $-\kappa^2$ as a bound state eigenvalue.  Our aim is to determine the single eigenvalue of the equation for every choice of $\beta$, which is the only parameter entering (\ref{aeq}). 
 
 Due to the singularity at the horizon we cannot integrate (\ref{aeq}) starting from the horizon.  However, we generated a 30th order Taylor series of $a$ around $\rho=0$, using the known series expansion coefficients of the metric components of the uniform black string ($f$, $g$, $h$). This allows us to calculate $a$ and $a'$ at $\rho=\Delta$, which is the same quantity which has also been used at the numerical integration of the equations of motion for the uniform black string. 
 
 We use the shooting method, starting from $\rho=\Delta$ to find the eigenvalue.  At each $\beta$ the coefficients of (\ref{aeq}) are calculated using the stored values of functions $f,\,g,$ and $h$ and their derivatives. We determine $\kappa$ at a large number of values of $\beta$ between the two singular points, 0 and 3.  The solid line of Fig. 3. represents the eigenvalue $\kappa=2\pi r_h/L$ as a function of $\beta$.  While at $\beta=0$ $\kappa$ reaches a finite value, $\kappa_c$, it diverges at $\beta=3$. This implies that for $\alpha<0$  black branes must undergo a Gregory-Laflamme transition before reaching the critical value and evaporate completely.  The limit of $\kappa$ as calculated from $\beta\not=0$ solutions agrees with $\kappa$ calculated from the critical solution. 
 \begin{figure}[htbp]
\begin{center}
\includegraphics[width=4in]{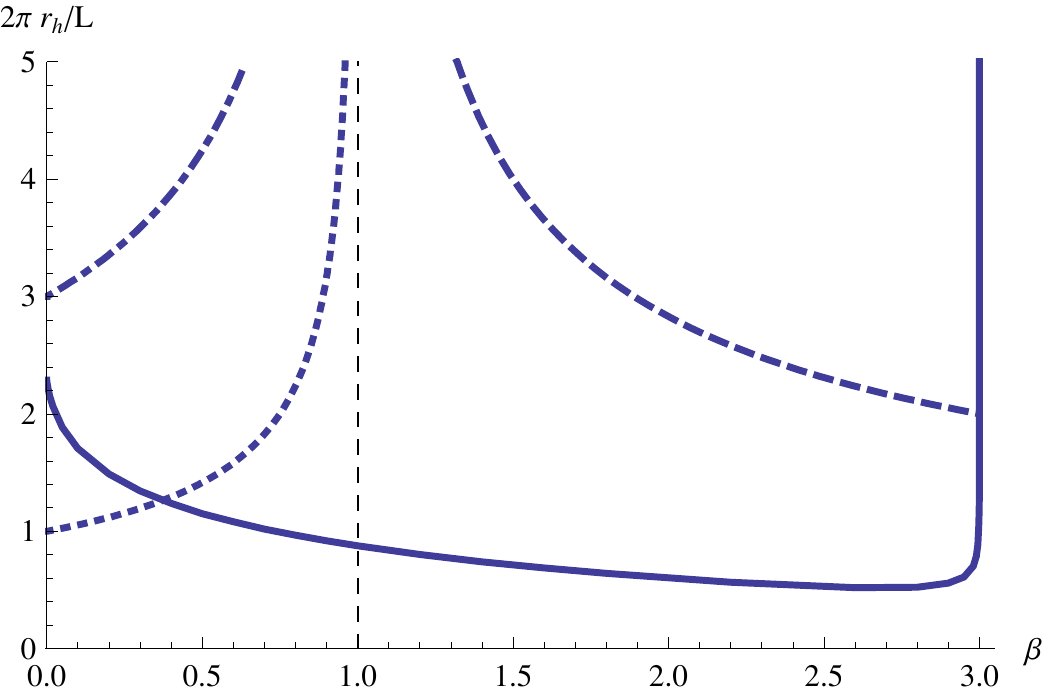}
\caption{Phase diagram of black strings. The solid line represents the first order transition line at $2\pi r_h/L=\kappa(\beta)$ as a function of $\beta$ between the two critical points, 0 and 3. The dotted, dash-dotted, and dashed lines represent time-lines of black strings at different values of $\alpha$} 
\label{Fig.3}
\end{center}
\end{figure}

 The time-lines of black strings are depicted by dash-dotted, dotted, and dashed lines in Fig. 3. These time-lines depend on $\alpha$ only. At fixed $\alpha$ different black strings may start their life at  different points but they follow the same line. If 
 \begin{equation}
r(\alpha,L)=\frac{2\pi \sqrt{8\alpha}}{L}>\kappa_c=2.2868
\end{equation}
then black strings completely avoid the instability transition before reaching the critical state (dash-dotted line in Fig.2). However, if $r(\alpha,L)<\kappa_c$ then every black string undergoes an instability transition and turns into a non-uniform black string before it could reach the critical line. The trajectory of black strings at such a value of $\alpha$ is represented by the dotted line.  Note that $r(\alpha,L)$ is completely determined by the geometry of space and the GB coupling. Furthermore, if $\alpha<0$ then $\beta>1$ and the time-line approaches $\beta=3$ (dashed line). Finally, if $\alpha=0$ then black strings follow the thin vertical dashed line at $\beta=1$.

\subsection{Non-uniform black strings near $\beta=0$}

In contrast to uniform black strings non-uniform black strings require the solution of a two variable problem. Therefore, it is quite a bit more complicated to investigate the fate of {\em non-unifom} black strings. However, we can substantially simplify matters if we restrict ourselves to sufficiently small values of $\beta$.  Though we cannot quite follow non-uniform black strings through the region of instability, we assume that if $\alpha>0$ they eventually also approach the $\beta=0$ line in the plot of Fig.2.  Of course, it is quite possible that nonuniform black strings also undergo an instability transition, before they could reach the critical state. Still the following calculations, pertaining to the behavior of nonuniform stings near $\beta=0$, certainly apply to a neighborhood of the critical point, $\beta=0$, $2\pi r_h/L\simeq \kappa_c$.

The simplest way to proceed is to investigate the horizon expansion (\ref{metric2}) with $w$-dependent coefficients. We write the following ansatz for the metric
\begin{equation}
ds^2=-\left(\rho+\frac{a_2}{2}\rho^2+...\right)dt^2+\frac{b_0(w)+b_1(w)\rho+...}{\rho+\frac{a_2}{2}\rho^2+...}dr^2+(\rho+1)^2(c_0(w)+c_1(w)\rho+...)d\Omega^2+(1+h_1(w)\rho+...)dw^2.
\label{metric3}
\end{equation}
Note that the $w$ dependence can be gauged away from $g_{tt}$ and from the zeroth order contribution to $g_{ww}$.   Expanding the equations of motion in $\rho$, the lowest order contribution to $G_{rw}$ implies that the periodic function $b_0(w)$ is independent of $w$.  In the next step one can solve the lowest order contributions to $G_{\theta\theta}=0$ and $G_{ww}=0$ equations for $b_1(w)$ and $h_1(w)$. Then the lowest order contributions to equations $G_{rr}=0$ and $G_{ww}=0$ are identical. They are second order equations for $c_1(w)$, containing $c_0(w)$ and its second derivative.   Solving the resulting equations for $c_1(w)$ the discriminant of the equation must be positive for a real solution. This discriminant has the form
\begin{eqnarray}
d&=&16 [c_0(w)]^4+(1-\beta)^2[c_0'(w)]^4-8(1-\beta)[c_0(w)]^3[2+c_0''(w)]-(1-\beta)^2c_0(w)[c_0'(w)]^2[2+3c_0''(w)]\nonumber\\
&+&2(1-\beta)[c_0(w)]^2\left\{4[c_0'(w)]^2+(1-\beta)c_0''(w)[2+c_0''(w)]\right\}.
\end{eqnarray}
Now we know that for uniform black strings $c_0(w)=1$ and for a small non-uniformity $c_0(w)$ must have the form
\begin{equation}
c_0(w)\simeq 1+ \lambda \cos(\kappa w).
\end{equation}
Then expanding $d$ in $\lambda$ and $\beta$ and keeping the leading order contributions only, we obtain
\begin{equation}
d\simeq 4[4\beta+(4+\kappa^2)\lambda \cos(\kappa w)].
\label{d2}
\end{equation}
(\ref{d2}) implies that the discriminant is positive for all $w$ only if 
\begin{equation}
\lambda<\frac{4\beta}{4+\kappa^2}.
\label{final}
\end{equation}
First of all (\ref{final}) implies that $\lambda\to 0$ when we approach the critical state, $\beta=0$. This makes our approximation of expanding $d$ in $\lambda$ correct for small $\beta$. Secondly, this implies that the non-uniform black string becomes a uniform black string again at the critical line. Thirdly, this implies the line of first order Gregory-Laflamme transition to the non-uniform state ends in a second order transition point at $\beta=0$. This is a well-known feature of first order transitions. 
 
\section{Conclusions}

We have studied uniform and non-uniform black strings in $D=5$ EGB gravity.  We paid particular attention to the singularities at the endpoints of range, $R$, $0\leq\beta<3$, where $\beta$ is defined in (\ref{beta}). No black string solutions exists outside $R$.   In particular, we found a critical solution at $\beta=0$ which, in contrast to solutions at $\beta>0$, has a horizon expansion that includes half-integer powers of the radial horizon variable, $\rho$ (\ref{rho}).  Not only the metric components, but all scalars, including the Ricci scalar and the Kretschmann scalar, are also singular (though finite) at $\beta=0$ (\ref{scalars}). 

At $\beta>0$ the metric components  are singular inside the horizon, at $\rho=-3\beta /7$ (\ref{singularity}). It is always possible to continue the metric to a region inside the horizon. Introducing Eddington-Fonkelstein coordinates
\begin{equation*}
u=t-\int d\rho\frac{\sqrt{g(\rho)}}{f(\rho)}
\end{equation*}
 the metric takes the form
\begin{equation*}
ds^2=-f(\rho)du^2+2\sqrt{g(\rho)}du\,d\rho+(\rho+1)^2d\Omega^2+h(\rho)dw^2,
\end{equation*}
regular everywhere including a neighborhood of the horizon.\footnote{Note that $g(\rho)>0$ on the interval $-3\beta/7<\rho<\infty$ for the whole range, $R$, of admissible values of $\beta$.} The singularity at $-3\beta/y$ moves to the horizon when $\beta\to0$. We have not been able to construct a metric inside this singular point. At $\beta=0$, the critical metric cannot be continued inside the horizon. This can be shown if near the horizon we transform the metric to Kruskal coordinates. We define the Kruskal coordinates for the critical string such that the coefficient of $dUdV$ is $-1$.  They can be obtained in a asymptotic expansion from the Eddington-Finkelstein coordinates $u$ and 
\begin{equation*}
v=t+\int d\rho\frac{\sqrt{g(\rho)}}{f(\rho)}.
\end{equation*}
We obtain
\begin{equation*}
ds^2=-dU\,dV+ (\rho+1)^2d\Omega^2+h[\rho]dw^2,
\end{equation*}
where $\rho$ has the following expression by the Kruskal coordinates $U$ and $V$
\begin{equation*}
\rho=-U\,V(1+c\, \sqrt{U\,V}+...),
\end{equation*}
where $c$ is a non-vanishing constant.  It is obvious we cannot cross the light like surfaces $U=0$ or $V=0$ which represent the horizon. We also showed that a geodesic of a massive probe falling toward the horizon reaches it in finite proper time (at $\tau=0$), but it cannot pass the horizon because its radial coordinate turns complex at $\tau>0$.
Therefore, the critical solution does not represent a black string. 

We used numerical techniques to calculate the solutions for the whole range $R$.  We calculated the ADM mass, tension, Hawking temperature, and entropy. These quantities are represented by form factors, depending on $\beta$ only, measuring the deviation from the $\alpha=0$ ($\beta=1$), Einstein gravity solutions. These form factors are defined in (\ref{hawk1})(\ref{m})(\ref{tau})(\ref{entscale}).They are plotted in Figs 1. and 2. While the Hawking-temperature tends to a finite value at $\beta=0$, it diverges at $\beta=3$.  The ADM mass and tension are modified only by a finite amount compared to pure Einstein gravity and bounded from above and below over the whole range, $R$. We gave a closed expression for the entropy in terms of an integral over the form factors, which was evaluated. The entropy is a monotonically decreasing function of $\beta$, vanishing in the limit $\beta=3$. This is not surprising in view of the divergence of the Hawking temperature. 

We paid  particular attention to the life line of black strings decaying via Hawking radiation.  We found that every black string follows a trajectory in the $(\beta, \kappa=2\pi r_h/L)$ plane, approaching (depending on the sign of $\alpha$) one of the singular lines in the $(\beta-r_h)$-plane. This is depicted in Fig. 3.  However, some of the black strings, depending whether $2\pi \sqrt{8\alpha}/L>\kappa_c=2.2868$, will encounter a line of first order singularities, which is the extension of the famed Gregory-Laflamme singularity.  This line is the solid line shown in the phase diagram of Fig. 3.  Under this line only non-uniform black string are stable.  As we commented before, non-uniform black strings may become unstable at a subsequent phase transition line.

Finally, we showed, by investigating non-uniform string solutions near the horizon that the amplitude of non-uniformity vanishes near the $\beta=0$ line.  This implies that non-uniform strings become uniform once again as they reach the $\beta=0$ line. As a corollary, we concluded that the line of first order phase transitions ends in a second order transition point at $\beta=0$. 

There are two unresolved problems concerning the fate of black strings. If $\alpha>0$ they are driven toward the critical state at $\beta=0$. That state, however is very problematic.  Though $g_{tt}$ has a linear zero and all scalars are finite at that zero, test particles that reach the $\rho=0$ point in finite time have nowhere to go. because the  $\rho(\tau)$ becomes complex as a function of proper time.  

As finding solutions for black holes in compactified spaces is even more difficult than finding black string solutions we have no way of comparing the entropy and relative stability of these solutions.  This is one of the reasons why we are not able to answer what happens with black strings that reach the end of their lives at the $\beta=0$ line.  Since the limit of their Hawking temperature is 
\begin{equation}
T_c=F_T(0)\frac{1}{4\pi \sqrt{8\alpha}}\simeq 0.85\frac{1}{4\pi \sqrt{8\alpha}},
\end{equation}
not zero, as a study based on $1/\rho$ expansions lead us to believe earlier, so they are certainly not frozen at that point. Their ADM mass is (\ref{m}) 
\begin{equation}
M_c=F_M(0)\frac{\sqrt{8\alpha}L}{2G}\simeq 1.23 \frac{\sqrt{8\alpha}L}{2G}.
\end{equation}
They cannot go into a black string state with a smaller ADM, mass, however, because such black string states do not exist. It seems that the only possibility is that some other, yet undiscovered, type of higher entropy static or possibly dynamic state exists, into which the black string can morph.  Another possible solution, at least for $D\geq6$, is that the inclusion of higher order Lovelock terms would cure this anomaly. The fate  of black stings at and beyond $\beta=0$ is one of the most important questions we would like to investigate in the future. Other questions include a possible numerical study of black holes in compactified spaces. 

This work used a Kaluza-Klein type compactification. It would be more difficult, but perhaps more interesting to repeat this work in a Randall-Sundrum type ADS space. 

\section{}
\begin{acknowledgments}
We thank Philip Argyres and Paul Esposito for discussions and Richard Gass for help in ''Mathematica''.. This work is supported in part by the DOE grant \# DE-FG02-84ER40153.
\end{acknowledgments}

\end{document}